\documentclass[aps,prl,amsmath,amssymb,reprint,superscriptaddress,preprintnumbers,showpacs,intlimits]{revtex4-1}
\usepackage{bm,latexsym,mathrsfs,enumerate,color}
\usepackage[mathcal]{euscript}
\usepackage[breaklinks=true,unicode=true,urlcolor = blue,colorlinks = true,citecolor = blue,linkcolor = blue]{hyperref}
\usepackage{graphicx}
\usepackage{subcaption}

\renewcommand{\vec}[1]{\bm{#1}}
\begin{document}


\title{Curvature induced chirality symmetry breaking in vortex core switching phenomena}

\author{Mykola I. Sloika}
     \email{sloika.m@gmail.com}
\affiliation{Taras Shevchenko National University of Kiev, 01601 Kiev, Ukraine}

\author{Volodymyr P. Kravchuk}
 \email{vkravchuk@bitp.kiev.ua}
 \affiliation{Bogolyubov Institute for Theoretical Physics, 03680 Kiev, Ukraine}

\author{Denis D. Sheka}
 \email{sheka@univ.net.ua}
\affiliation{Taras Shevchenko National University of Kiev, 01601 Kiev, Ukraine}

\author{Yuri~Gaididei}
 \email{ybg@bitp.kiev.ua}
 \affiliation{Bogolyubov Institute for Theoretical Physics, 03680 Kiev, Ukraine}

\date{\today}

%
%

\begin{abstract}
The interplay between magnetic vortex polarity, chirality and the curvature of the underlying surface results in a dependence of the vortex polarity switching efficiency on the vortex chirality. The switching is studied numerically by applying a short Gauss pulse of the external magnetic field to a spherical cap within its cut plane. The minimum field intensity required for the switching essentially depends on the vortex chirality and it does not depend on the initial vortex polarity. This effect decreases with the curvature radius increasing and it vanishes in the planar limit.
\end{abstract}


%
%

%
%


\maketitle
Sub-micron sized ferromagnetic nanoparticles are of high applied interest now. During the last decade the main attention was paid to planar nanostructures such as magnetic nanodisks, nanorings etc., where the competition of short-range exchange and long-range dipole-dipole interactions  results in a vortex magnetic ground state \cite{Hubert98}. Magnetic vortex has an in-plane magnetization circulation around small vortex core where the magnetization is perpendicular to the magnet plane. Thus the vortex is characterized by two discrete indices: \emph{chirality} (counterclockwise, $c = +1$, or clockwise, $c = -1$), the sense of magnetization circulation, and \emph{polarity} (up, $p = +1$ or down, $p = -1$), the sense of the core magnetization direction. A nanoparticle with its vortex ground state is usually considered as a carrier of one or two bits of information, which is a promising feature for fabrication of nonvolatile memory devices \cite{Bohlens08,Pigeau10,Yu11a,Hertel13}. For this purpose the possibility of chirality and polarity control is of principle. The most simple and practically reliable way of the vortex polarity switching is to apply a short in-plane pulse of magnetic field \cite{Waeyenberge06,Hertel07}.

Recently it was demonstrated both theoretically \cite{Sheka13b} and experimentally \cite{Streubel12,Streubel12a} that the vortex state can be ground one for hemispherical magnetic nanocaps. In this Letter we study the vortex polarity switching in spherical caps of different curvature radii. The curvature is shown to break the chiral symmetry of the vortex polarity switching process.

We base our study on the numerical micromagnetic simulations \cite{magpar,Scholz03a} of Permalloy \footnote{Material parameters are the following: exchange constant $A = 1.05\times 10^{-11}$ J/m, saturation magnetization $M_s = 7.96\times 10^5$ A/m, damping constant $\alpha = 0.01$ and on--site anisotropy was neglected. The average mesh size was 2.5 nm for all samples.} samples.

\begin{figure}
\includegraphics [clip,width=0.9\columnwidth, angle=0]{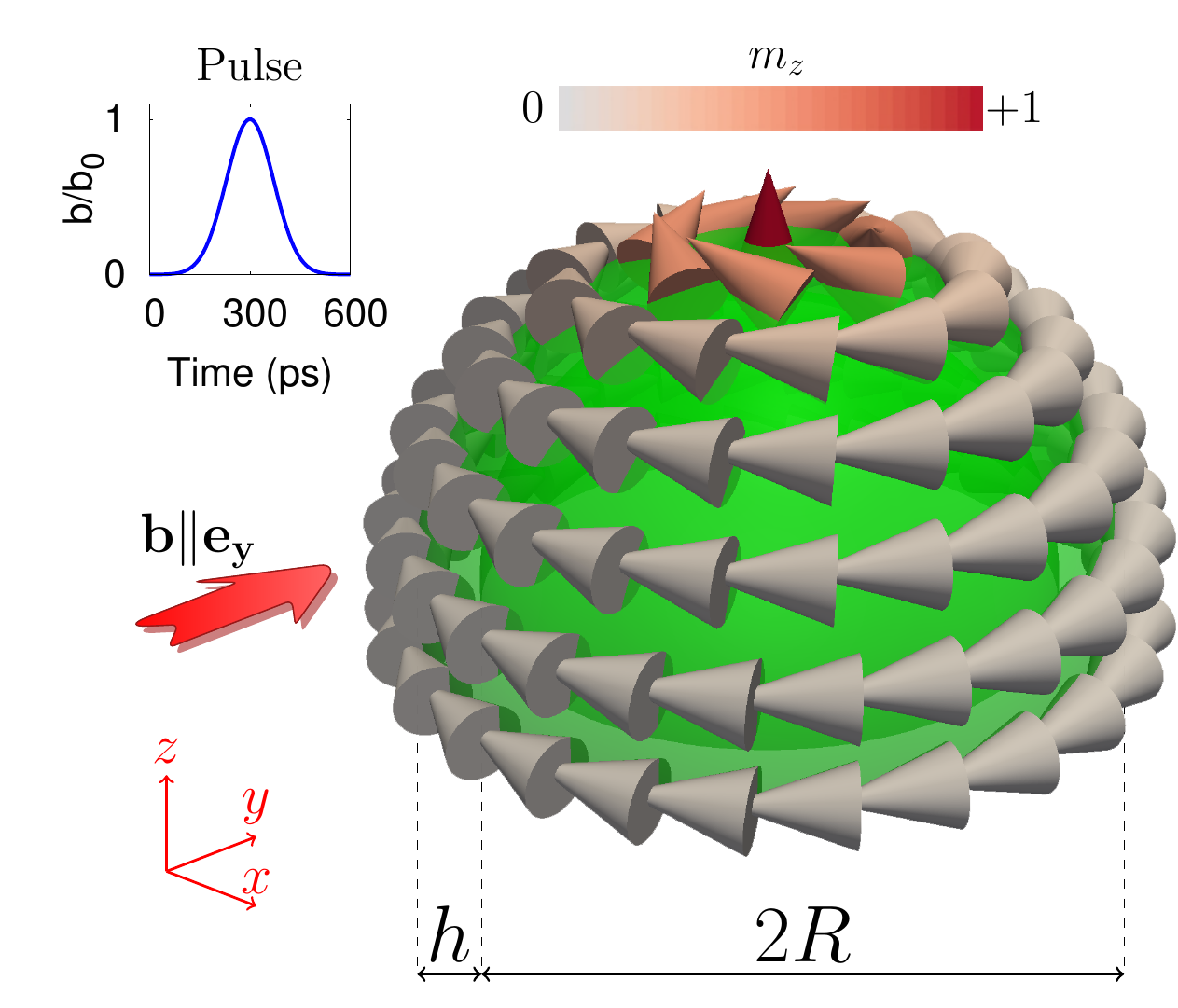}
\caption{(Color online) Vortex state of the hemispherical magnetic cap (right). Magnetization distribution is shown by cones. Large arrow indicates the direction of magnetic field pulse $\vec{b}||\vec{e}_y$ of Gauss time profile (left).}
\label{fig:VortexOnCup}
\end{figure}

We start from the hemispherical cap with inner radius $R=50$ nm and thickness in radial direction $h=10$ nm. Notations and geometrical details of the problem are shown in the Fig.~\ref{fig:VortexOnCup}. According to the phase diagram of equilibrium states of soft magnetic hemispherical shells~\cite{Sheka13b}, the used geometrical parameters correspond to the vortex ground state of the cap. This ground state was obtained using the energy minimization procedure, see Fig.~\ref{fig:VortexOnCup}. Then, in the same way as it was done for planar disks \cite{Hertel07}, we apply a time dependent and spatially uniform pulse of magnetic field $\vec{b}(t)$ in $y$-direction (perpendicularly to the symmetry axis $\vec e_z$). As in Ref.~\onlinecite{Hertel07} the time profile of the pulse was chosen to be Gaussian one:
 $\vec{b} = \vec{e}_y b_{0} \exp\left[-(t - 3 \tau)^2/\tau^2\right]$,
where $t$ denotes time and $\tau = 100$~ps is the pulse duration. Using numerical integration of the Landau-Lifshitz-Gilbert (LLG) equation the pulse induced vortex dynamics is simulated. Similarly to the case of disk~\cite{Hertel07} the vortex behavior depends on the pulse amplitude $b_0$. If the amplitude $b_0$ exceeds some critical value $b_c$ the vortex polarity $p$ switches to the opposite value via creation of an intermediate vortex-antivortex pair with subsequent annihilation of the new-born antivortex with the original vortex. Otherwise, the vortex core relaxes to the cap center without polarity switching.

\begin{figure}
\includegraphics[width=\columnwidth]{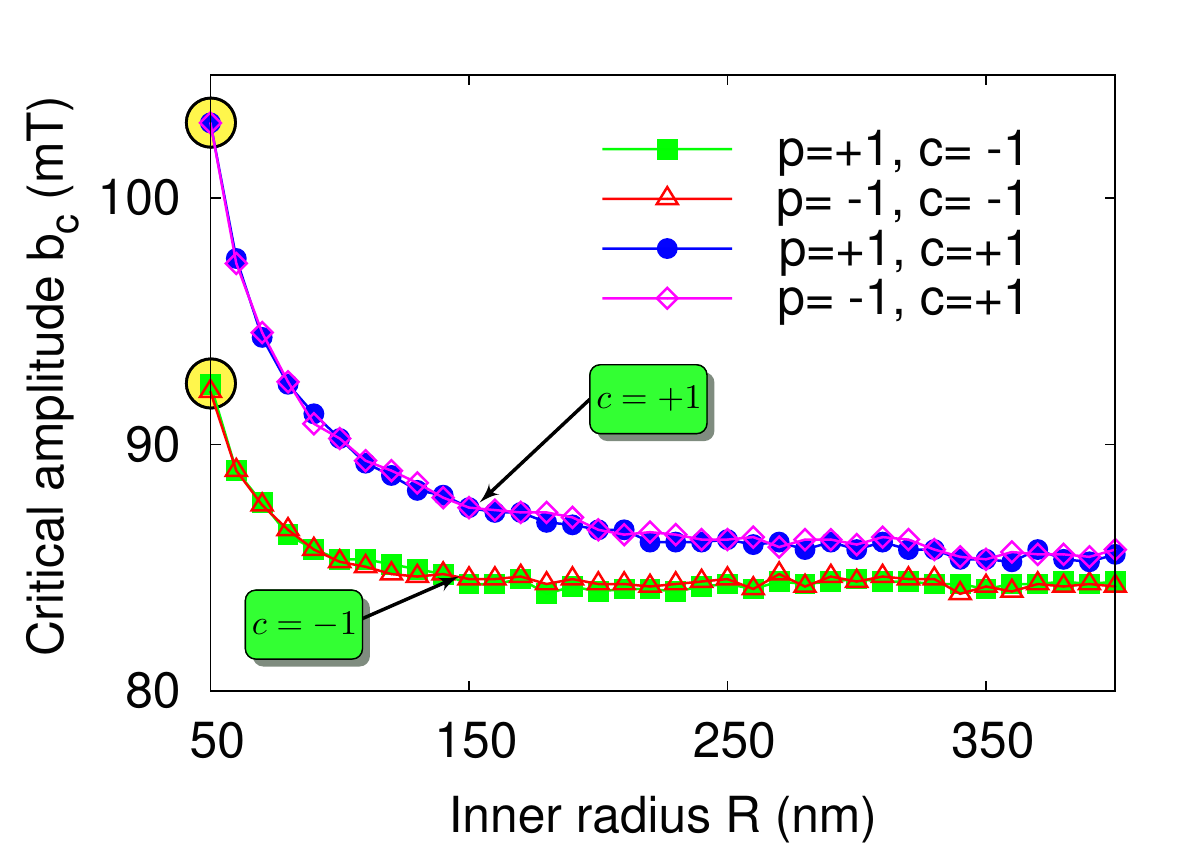}
\caption{(Color online) Switching amplitude of the field pulse for all possible polarity-chirality combinations as a function of the curvature radius of the cap. The cycled symbols correspond to the hemispherical shell with $R=50$~nm.}
\label{fig:HextVsRadius}
\end{figure}

We determine the critical amplitude $b_c$ for all four possible combinations of chiralities and initial polarities. We find out that $b_c$ does not depend of the polarity $p$ but it does depend on the chirality $c$, see the circled points in Fig.~\ref{fig:HextVsRadius}. One can see that for the case $c = +1$ the critical amplitude $b_c$ is greater by 12\% than for the case $c = -1$.

\begin{figure}
\includegraphics[width=0.8\columnwidth]{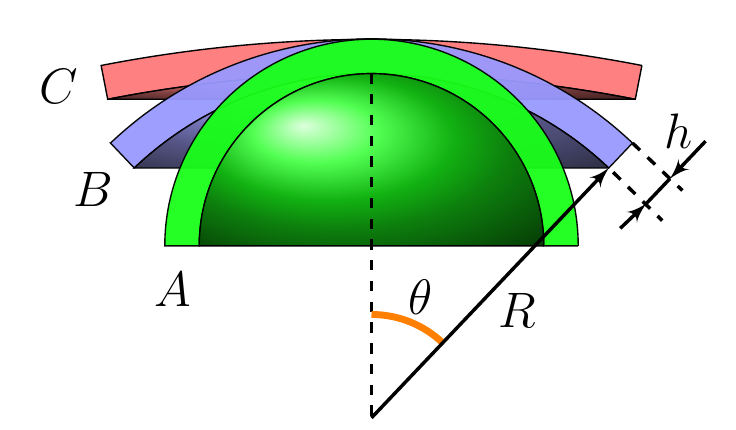}
\caption{(Color online) Continuous transition from a hemispherical cap to a planar disk: (A) -- the hemispherical shell with inner radius $R = 50$~nm,  (B) -- $R = 100$~nm, (C) -- $R = 400$~nm. Thickness $h = 10$~nm and volume are kept constant.}
\label{fig:fromCupToDisc}
\end{figure}

Since in planar structures the polarity switching process is symmetrical with respect to the vortex chirality we study role of the curvature by carrying out the described numerical experiment for spherical caps with different curvature radii. Keeping the sample volume $V$ and thickness $h$ constant, we increase the inner radius $R$ with the corresponding decreasing of the cutoff angle $\theta$, see Fig.~\ref{fig:fromCupToDisc}. Thus the continuous transition from the hemispherical cap to planar disk ($R\to\infty$) is realized.

For each of the spherical caps the switching amplitude $b_c$ was determined for all possible polarity-chirality combinations. The results are presented in the Fig.~\ref{fig:HextVsRadius}. As one can see, the vortex with clockwise magnetization circulation ($c=-1$) always requires lower pulse amplitude for switching. Moreover, the difference between switching amplitudes for opposite chiralities vanishes with the curvature radius increasing. Therefore in the planar limit the previous results are reproduced. Also it is important to note that the switching amplitude $b_c$ does not depend on the initial vortex polarity.

\begin{figure*}[t]
\includegraphics[width=0.8\textwidth]{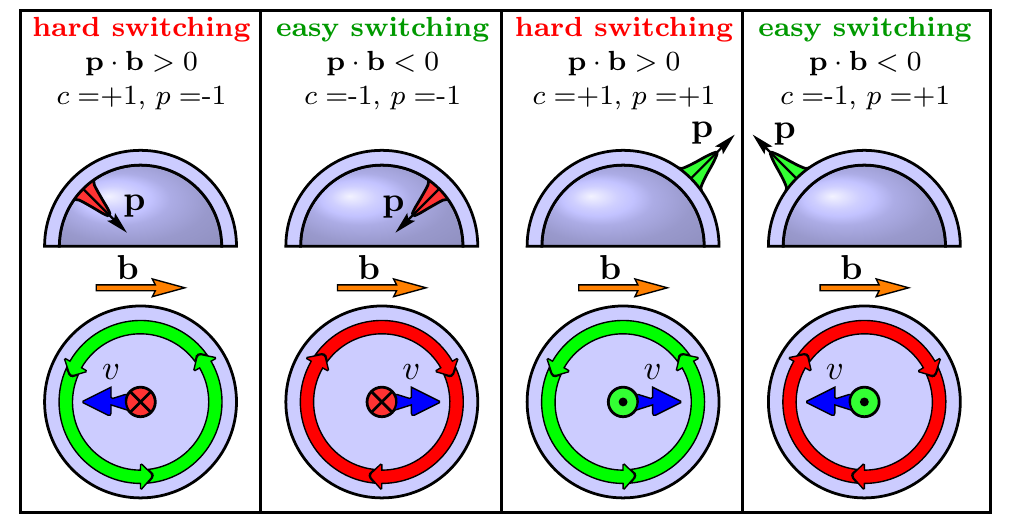}
\caption{(Color online) Qualitative explanation of the chirality symmetry breaking effect in the process of the vortex core reversal. Side cross-section and top view are shown for all possible polarity-chirality combinations. Vector $\vec v$ denotes the initial vortex core velocity when the pulse $\vec b$ is applied. The sign of the product $\vec p\cdot\vec b$ always coincides with sign of chirality $c$, that breaks the chirality symmetry in the switching process.}
\label{fig:Effect}
\end{figure*}
To provide the qualitative explanation of the observed phenomenon we first refer to recent studies [Refs.~\onlinecite{Khvalkovskiy10,Yoo10}] where it was shown that the critical vortex core velocity~\cite{Lee08c}, which is required for the polarity switching, depends on the perpendicular component of the magnetic field. Namely, if the external field direction coincides with the vortex polarity ($\vec b\cdot\vec p>0$) the switching is ``hard'': the critical velocity is higher and consequently the more strong excitation is needed; and vice-versa, for the case $\vec b\cdot\vec p<0$ the switching is ``easy'', critical velocity is lower and so the less intensive excitation is needed. Here the polarity vector $\vec p$ is the unit magnetization vector in the vortex center. In the case of an in-plane field applied to the planar structure one always has $\vec b\cdot\vec p=0$.  However, it is not a case for spherical shells. Since $\vec p$ remains normal to the surface the sign of the product $\vec b\cdot\vec p$ is not constant anymore but depends on the direction of shifting with respect to the field vector $\vec b$.

It is well known~\cite{Mertens00} that the vortex core shifted from the equilibrium position in a nanodot gyrates around the equilibrium point. For the studied samples the period of the vortex gyration $\sim10^{-9}$ s is an order of magnitude greater than the typical switching time $\sim10^{-10}$ s, the time, which elapses from the pulse beginning to the switching moment. Thus, only direction in which the vortex core starts to move matters for the switching process. This direction can be easily determined using Thiele equation~\cite{Thiele73,Huber82a}. In the case of small vortex core displacements similarly to the flat geometry one can write down this equation in the form of the force balance $G\left[\dot{\vec{\mathcal{R}}}\times\vec e_z\right]
+\vec{F}=0$.
Here $\vec{\mathcal{R}}=(\mathcal{X},\mathcal{Y})$ gives the position of the vortex core, $G\propto p$ is a gyroconstant, and $\vec{F}=-\mathrm{d}\mathcal{E}/\mathrm{d}\vec{\mathcal{R}}$ is the external force with $\mathcal{E}(\mathcal{\vec{\mathcal{R}}})$ being the total energy of the system. For magnetically soft samples the total energy has three contributions $\mathcal{E}=\mathcal{E}_{ex}+\mathcal{E}_{ms}+\mathcal{E}_{z}$, where the exchange term $\mathcal{E}_{ex}\sim\mathcal{R}^2$ as well as magnetostatic term $\mathcal{E}_{ms}\sim\mathcal{R}^2$ quadratically depend on vortex displacement~\cite{Guslienko01a,Guslienko02a,Gaididei10} while the interaction energy with the external magnetic field $\mathcal{E}_z$ is linear with respect of the vortex displacement   $\mathcal{E}_z\sim\mathcal{R}$~\cite{Guslienko01a,Guslienko02a}. Therefore for the small vortex displacements the Zeeman
contribution dominates and $\vec F\approx\vec
-\mathrm{d}\mathcal{E}_z/\mathrm{d}\vec{\mathcal{R}}$. The Zeeman interaction tends to increase an area magnetized
along the applied field and the force $\vec F$ becomes chirality sensitive. A simple calculation shows that $\vec{F}\propto -c[\vec b\times\vec e_z]$ for  the case $\mathcal{R}\ll R$. Substitution $\vec{F}$ into Thiele equation enables one to obtain the \emph{initial} direction of the vortex core motion when the field pulse is applied $\dot{\vec{\mathcal{R}}}\propto cp\vec b$. This is illustrated in the Fig.~\ref{fig:Effect} for all possible polarity-chiraliry combinations.
As one can see the sign of the product $\vec p\cdot\vec b$ always coincides with the sign of chirality $c$, and therefore the vortex with chirality $c=-1$ requires less excitation amplitude for the polarity switching as compared with $c=+1$. This is the essence of the chirality symmetry breaking in the switching process.

\begin{figure*}[t]
\begin{subfigure}{.49\textwidth}
 \includegraphics[width=\columnwidth]{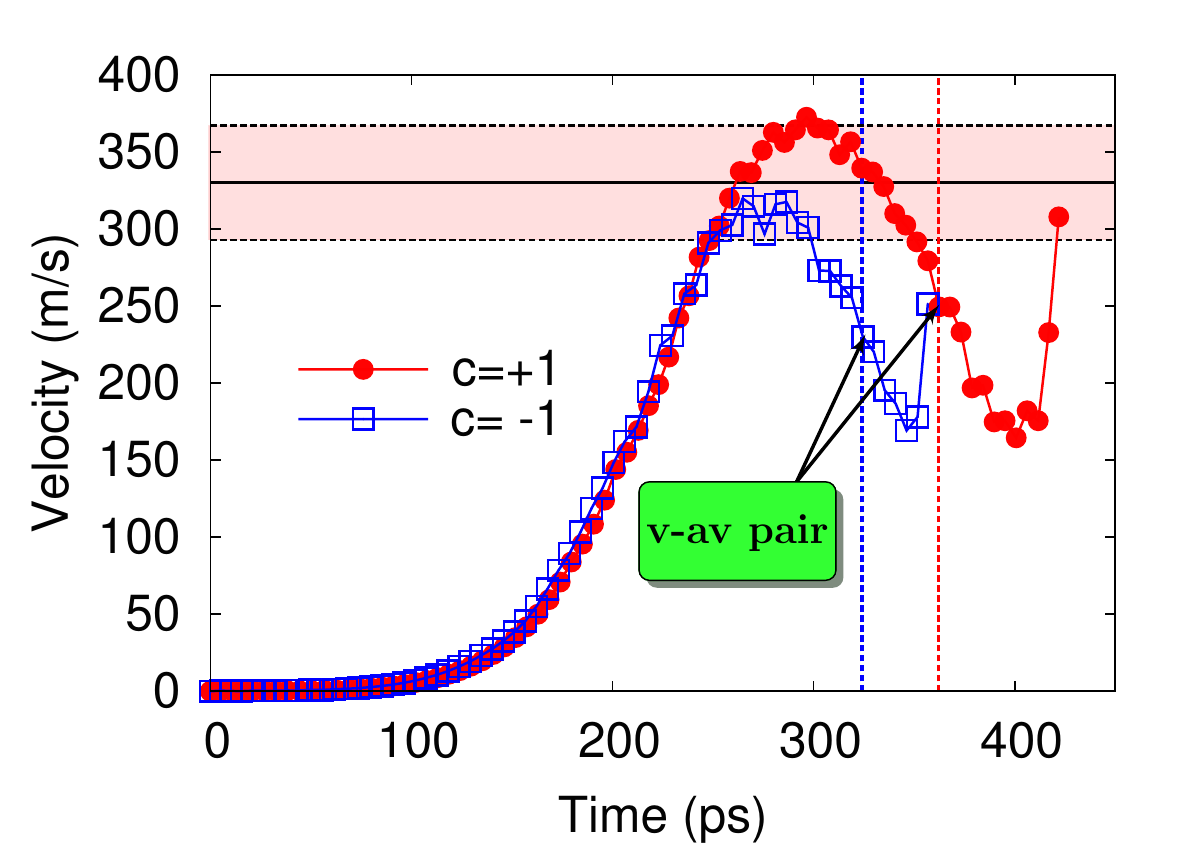}
\caption{$R = 50 $ nm}
\label{fig:VelocityVsTimeCaps}%
\end{subfigure}
\begin{subfigure}{.49\textwidth}
\includegraphics[width=\columnwidth]{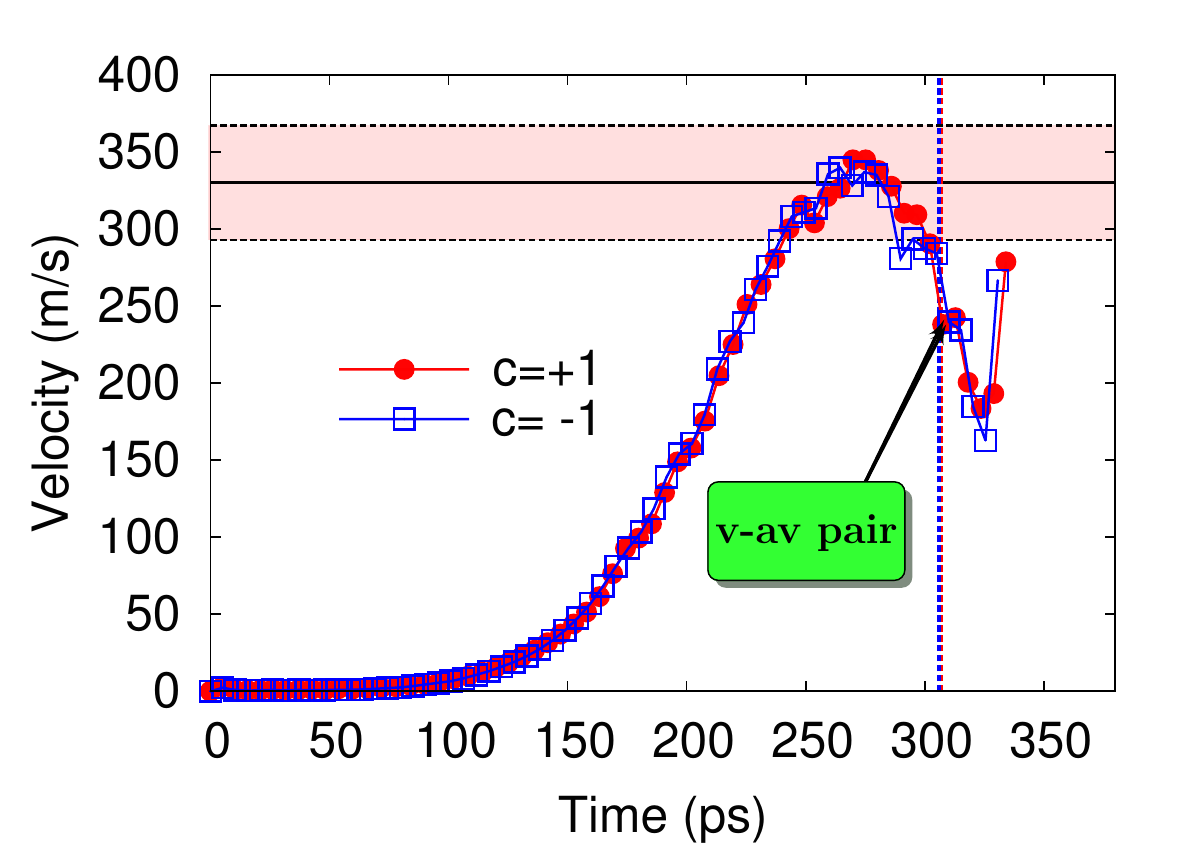}
\caption{$R = 400 $ nm}
\label{fig:VelocityVsTimeDisk}
\end{subfigure}
\caption{(Color online) Temporal evolution of the vortex core velocity for different chiralities and curvature radii. In all cases the pulse amplitude is the same $b_0=103.1$~mT. Solid horizontal line corresponds to the critical velocity  for Permalloy disk-shaped nanoparticle $330$~m/s with error-lines $\pm 37$ [\onlinecite{Lee08c}]. Dashed vertical lines as well as callouts point to the time moment when the vortex-antivortex pair appears. The dependencies are terminated by the annihilation moment.}\label{fig:velocities}
\end{figure*}

To verify the proposed qualitative explanation of the effect we consider how the vortex core velocity $v(t)$ evolves in time, see Fig.~\ref{fig:velocities}. On the whole, the vortex reversal process in spherical caps and planar disks are similar~\cite{Lee08c,Hertel07}: when the field is switched on, the vortex accelerates, and near the vortex core an area with the magnetization component opposite to the vortex polarity (so called ``dip'') is formed. Initially the dip amplitude is very small. However, when the core velocity reaches some critical value (maximum value on the dependencies $v(t)$, see Fig.~\ref{fig:velocities}) the dip amplitude rapidly increases, and consequently the vortex-antivortex pair with polarities opposite to the initial vortex polarity is born (see horizontal dashed lines in the Fig.~\ref{fig:velocities}). Then annihilation of the
initial vortex with the newly born antivortex finalizes the switching process. It is significant for hemispherical shells that for the vortex shifted from its pole position the projection of  the core magnetization on the external field direction is finite and in this way, the dip creation can be either facilitated or suppressed. For example in the case $c=-1$ one always has $\vec b\cdot\vec p<0$ in the initial stage of the vortex movement. That means that there is a field component which favours the dip creation, and therefore the critical velocity is smaller~\cite{Khvalkovskiy10,Yoo10} compared with the opposite case $c=+1$, see Fig.\ref{fig:VelocityVsTimeCaps}. With the curvature radius increasing the described effect vanishes, see Fig.~\ref{fig:VelocityVsTimeDisk}.

To summarize, for the spherical magnetic caps we have studied the vortex polarity switching process by a magnetic field pulse applied in the cut plane. We demonstrate that due to the curvature this process loses the chiral symmetry typical for planar systems: the clockwise chirality requires lower magnetic pulse to switch the vortex polarity than the counterclockwise chirality. Main reason is appearance of the normal component of external magnetic field at the core of shifted vortex. Depending on the chirality this can suppress or favour the switching process.

%

%

\end{document}